\documentclass[12pt]{article}
\usepackage{scicite}
\usepackage{times}
\newenvironment{sciabstract}{%
\begin{quote} \bf}
{\end{quote}}

\newcounter{lastnote}
\newenvironment{scilastnote}{%
\setcounter{lastnote}{\value{enumiv}}%
\addtocounter{lastnote}{+1}%
\begin{list}%
{\setlength{\leftmargin}{.22in}}
{\setlength{\labelsep}{.5em}}}
{\end{list}}

\newcommand{\unit}[1]{\, \mathrm{#1}} 
\newcommand{\degr}{\!\!\!\:^{\circ}}
\newcommand{\degree}{^{\circ}\!}

\usepackage{graphicx}

\title{A Uranian Trojan and the frequency of temporary giant-planet co-orbitals.} 

\author
{Mike Alexandersen,$^{1\ast}$ Brett Gladman,$^{1}$ Sarah Greenstreet,$^{1}$\\
J.J. Kavelaars,$^{2}$ Jean-Marc Petit,$^{3}$ Stephen Gwyn$^{2}$ \\
\\
\normalsize{$^{1}$Department of Physics and Astronomy, University of British Columbia,}\\
\normalsize{6224 Agricultural Road, Vancouver, BC V6T 1Z1, Canada}\\
\normalsize{$^{2}$National Research Council of Canada, Victoria, BC V9E 2E7, Canada}\\
\normalsize{$^{3}$Institut UTINAM, CNRS-UMR 6213, Observatoire de Besan\c{c}on,}\\
\normalsize{BP 1615, 25010 Besan\c{c}on Cedex, France}\\
\\
\normalsize{$^\ast$E-mail:  mikea@astro.ubc.ca}
}
\date{}

\begin{document} 
\baselineskip24pt
\maketitle 

\begin{sciabstract}

  Trojan objects share a planet's orbit, never straying far from the triangular Lagrangian points, $60\degree$ ahead of (L4) or behind (L5) the planet.
We report the detection of a Uranian Trojan; 
in our numerical integrations, 2011 QF$_{99}$ oscillates around the Uranian L4 Lagrange point for $>70$~kyr and remains co-orbital for $\sim1$~Myr before becoming a Centaur.
  We constructed a Centaur model, supplied from the transneptunian region, to estimate temporary co-orbital capture frequency and duration (to factor of two accuracy), finding that at any time $0.4\%$ and $2.8\%$ of the population will be Uranian and Neptunian co-orbitals, respectively.
  The co-orbital fraction ($\sim2.4\%$) among Centaurs in the IAU Minor Planet Centre database is thus as expected under transneptunian supply.
\end{sciabstract}

During 2011 and 2012, we used the Canada-France-Hawaii Telescope to perform a 20-square-degree survey designed to detect Trans-Neptunian Objects (TNOs) and objects between the giant planets with apparent r-band magnitude $m_r<24.5$, and track all detections for up to 17 months. 
The project was accurately calibrated\cite{SM} in order to constrain the size and orbital parameter distributions of populations resonant with Neptune.
Constraining the distribution of these populations is essential, as they in turn set constraints on models of the evolution of the outer Solar System. 

As part of this survey, we detected 2011 QF$_{99}$\cite{alexandersen13a} at a heliocentric distance of $20.3$ AU, where its apparent magnitude $m_r=22.6\pm0.1$ sets its absolute magnitude at $H_r=9.6$ ($H_g=10.3$ assuming a typical colour $g-r\approx0.7$). 
This magnitude indicates that 2011 QF$_{99}$ is $\sim60\unit{km}$ in diameter, assuming a $5\%$ albedo. 
As more observations constrained the orbit, it became clear that 2011 QF$_{99}$ was not simply a Centaur that happened to be near the distance of Uranus.
Our current astrometry, consisting of 29 measurements from 7 dark runs with total arc of 419 days, indicates the following orbital elements:
$a=19.090\pm0.004\unit{AU}$, $e=0.1765\pm0.0007$, $i=10.\degr811\pm0.\degr001$, $\Omega=222.\degr498\pm0.\degr001$, $\omega=287.\degr51\pm0.\degr11$, $T=246\,4388\pm11\unit{JD}$.
Here $a$, $e$, $i$, $\Omega$, $\omega$, $T$ are the osculating J2000 barycentric semi-major axis, eccentricity, inclination, longitude of ascending node, argument of pericentre and Julian day of pericentre. 
The low eccentricity along with a semi-major axis similar to that of Uranus ($a_{U}\approx19.2\unit{AU}$) indicated that 2011 QF$_{99}$ might be a Uranian co-orbital. 
Co-orbital bodies are in the 1:1 mean-motion resonance with a planet (thus having the same orbital period) and a librating (oscillating) resonant angle $\phi_{11}=\lambda-\lambda_{Planet}$. 
Here $\lambda$ is the mean longitude, which is the sum of $\Omega$, $\omega$ and the mean anomaly. 
$\phi_{11}$ roughly measures how far ahead in orbital phase the object is relative to the planet. 
For co-orbital motion, $\phi_{11}$ librates around one of four values\cite{mikkola06}. 
Quasi-satellites librate around $0\degree$; in the co-orbital frame these move like retrograde satellites, despite being outside the planet's gravitational dominance. 
Leading and trailing Trojans librate around L4 ($60\degree$ ahead of planet) and L5 ($300\degree$ ahead $=60\degree$ behind planet), respectively. 
Horseshoe orbits execute librations around L3 ($180\degree$ from the planet) with high-amplitudes that encompasses the L3, L4 and L5 Lagrange points. 

A short ($50\unit{kyr}$) numerical integration showed that 2011 QF$_{99}$ is librating around the leading (L4) Lagrange point (Fig. 1).
Could 2011 QF$_{99}$ be a primordial Trojan?
Jupiter hosts a large population of Trojan asteroids stable for $5\unit{Gyr}$. 
The recently-detected stable population of Neptunian Trojans are now believed to outnumber Jovians for objects with radius $>50\unit{km}$\cite{chianglithwick05,sheppardtrujillo10b}.
In contrast, the Trojan regions of Saturn and Uranus are believed to be mostly unstable\cite{nesvornydones02,dvorak10} and are unlikely to host long-lived Trojans, 
although a few stable niches exist\cite{dvorak10}; it is unclear how migration affects the likelihood of these niches being populated\cite{kortenkamp04,kortenkampjoseph11}. 

In longer integrations, 
using the $10\unit{Myr}$ time scale typically used to determine the dynamical class of outer Solar System objects \cite{gladman08,lykawkamukai07}, 
both the nominal orbit of 2011 QF$_{99}$ and all other orbits within the (already small) orbital uncertainties librate around the L4 Lagrange point for at least the next $70\unit{kyr}$ (Fig. 2 left column, and Fig. S1).
On time scales of $100\unit{kyr}$ to $1\unit{Myr}$, all integrated orbits transition out of the L4 Trojan region\cite{SM}, 
either escaping directly to scattering behaviour (that is, become Centaurs) or transitioning to other co-orbital behaviour before escaping and scattering away within $3\unit{Myr}$.

We considered the possibility that the initial investigation missed a small phase-space niche, stable for $4\unit{Gyr}$ or that systematic errors could result in the real orbit being offset by several tens times
the nominal uncertainties.
We thus integrated $10^5$ test particles filling the region within
$\Delta a=\pm0.1\unit{AU}$, $\Delta e=\pm0.004$, and
$\Delta i=\pm0.\degr2$ of the nominal orbit for up to $0.1\unit{Gyr}$,
until they crossed the orbit of Saturn or Neptune.
All 100,000 particles were eliminated within $100\unit{Myr}$,
most within the usual $10\unit{Myr}$ stability of Centaurs\cite{dones99,tiscarenomalhotra03}.
This rejects the idea that 2011 QF$_{99}$ has been a Uranian
Trojan for very long; it must instead be a Centaur recently
temporarily trapped into L4 libration.

Temporary co-orbitals are known elsewhere in the Solar System\cite{SM}.
In this survey, 2011 QF$_{99}$ was the only object with a semi-major axis within the planetary region (defined here as $a<34\unit{AU}$ to include Neptune co-orbitals but exclude the exterior stable transneptunian populations). 
The Canada-France Ecliptic Plane Survey (CFEPS) detected three $a<34\unit{AU}$ objects and the IAU Minor Planet Centre (MPC) database contains 247 objects with $6\unit{AU}<a<34\unit{AU}$ as of 2013-Jul-09.
We seek to determine if (to factor of three) it is reasonable that, in a model of Centaur supply from the
Scattering Disk, a large-enough fraction should be in resonance at any time
to explain the observed discovery of 2011 QF$_{99}$ and other temporary co-orbitals. 
To address this question, we estimated the fraction of Centaurs in temporary co-orbital states with Uranus and Neptune, similar to
what has been done for the Earth\cite{moraismorbidelli02} and Venus\cite{moraismorbidelli06}. 
Even though the scattering population is depleting with time, the co-orbital fraction does not\cite{SM}.
Here, ``scattering objects'' are those\cite{gladman08,petit11} that experience $\Delta a>1.5\unit{AU}$ in $10\unit{Myr}$; scattering objects with $a<30\unit{AU}$ are called ``Centaurs'', whereas those with $a>30\unit{AU}$ are the ``Scattering Disk''. 

Using a model of the orbital distribution\cite{kaib11} of today's scattering TNOs\cite{SM}, we simulated the interactions of scattering objects with the giant planets over $1\unit{Gyr}$, building a relative orbital distribution for the $a<34\unit{AU}$ region\cite{SM}.
The simulation outputs the state vector of planets and all $a<34\unit{AU}$ particles at $300\unit{yr}$ intervals. 
This output interval was chosen so that the few-kyr variation of the resonant argument $\phi_{11}$ would be well sampled (see Fig. 2), allowing detection of short-term co-orbitals of the giant planets. 
Such a meticulous search for co-orbitals trapped from an armada of incoming scattering objects is essential in order to accurately estimate the trapping fraction. 
An earlier analysis\cite{hornerevans06} started with a sample of currently known Centaurs, which was biased towards the lowest-$a$ Centaurs by observational selection, resulting in much lower trapping rates for Uranus and Neptune than we find. 

Simulated scattering objects predominantly entered the giant planet region ($a<34\unit{AU}$) at intermediate inclinations and eccentricities, as has been previously shown\cite{levisonduncan97,tiscarenomalhotra03}. 
After analysing the particle histories to find co-orbital trapping\cite{SM}, we find that $0.4\%$ and $2.8\%$ of the $a<34\unit{AU}$ population is, at any time, in co-orbital motion with Uranus and Neptune, respectively (with less than factor 2 variation\cite{SM}). 
This $3.2\%$ fraction is much larger than the $\approx0.1\%$ of near-Earth asteroids temporarily trapped in Earth and Venus co-orbital motion \cite{moraismorbidelli02,moraismorbidelli06}, presumably due to the fractionally larger co-orbital regions of the giant planets. 
We find that the simulated Uranian and Neptunian co-orbitals consisted of, respectively, $64\%$ and $54\%$ in horseshoe orbits, $10\%$ and $10\%$ quasi-satellites\cite{SM} and $26\%$ and $36\%$ Trojans, equally distributed between the L4 and L5 clouds. 
The duration of Uranian co-orbital captures in our simulation had mean, median and maximum values of $108\unit{kyr}$, $56\unit{kyr}$ and $2.6\unit{Myr}$, respectively, and $78\unit{kyr}$, $46\unit{kyr}$ and $18.2\unit{Myr}$, respectively, for Neptune.

To explore the strength of observational biases, we use the CFEPS Survey Simulator\cite{gladman12}, expanded to include our additional coverage. 
The survey simulator applies observational biases to the population model (Fig. 3 top) from the dynamical simulation\cite{SM} to simulate what a survey would observe (Fig. 3 bottom).
The absolute $H$-magnitude distribution (a proxy for the size distribution) of objects is important when modelling flux limited surveys. 
Our first attempt to use a single exponential distribution ($\mathrm{d}N/\mathrm{d}H\propto10^{\alpha H}$\cite{SM}) with $\alpha\approx0.8$, as measured for $H_g<9.0$ TNOs\cite{elliot05,petit11} and Neptunian Trojans\cite{sheppardtrujillo10b}, was rejected at a high level of confidence, predicting that small ($H_g>11.0$) objects should account for $81\%$ of the detections; our observations have no such objects. 
The $H$-mag distribution of Neptunian Trojans cannot continue as a single exponential\cite{sheppardtrujillo10b} beyond $m_R\approx23.5$ ($H_g\approx9.3$ assuming $g-R=0.5$ and typical distance of $30\unit{AU}$). 
The scattering objects also reject a single exponential and are better represented by a ``divot'' $H$-mag distribution\cite{shankman13}, where the number density drops at $H_g\approx9.0$ by a contrast factor (ratio of density just before and after the divot) of 6, then continues with a second shallower exponential with $\alpha\approx0.5$. 
Although we lack the statistics to independently constrain a divot, simply adopting the above parameters provides better agreement between simulated and observed populations (see Fig. 3), with small ($H_g>11.0$) objects only providing $23\%$ of the simulated detections. 

Simulating our available calibrated fields we find that $0.9\%$ of the $a<34\unit{AU}$ detections should be Uranian co-orbitals and $2.2\%$ should be Neptunian co-orbitals (Fig. 3).
Thus, because of where we looked and the survey limits, detection of Uranian co-orbitals was enhanced compared to the intrinsic fraction ($0.4\%$), whereas the Neptunian co-orbitals were slightly biased against (relative to $2.8\%$).
Of the 247 objects with $6\unit{AU}<a<34\unit{AU}$ currently in the MPC database, around six (including 2011 QF$_{99}$) have been identified as temporary co-orbitals of Uranus and Neptune\cite{SM}, yielding $\sim2.4\%$. 
This is within a factor of two of our $3.2\%$ model prediction, although the unknown pointing history and survey depths make detailed modelling impossible. 
Because our simulations show that no large (not even factor of two) biases exist towards or against detecting co-orbitals, the MPC co-orbital fraction may be close (within a factor of a few) to the intrinsic fraction.
Thus 2011 QF$_{99}$ is a Uranian Trojan that is part of a roughly constant population of transient co-orbitals, temporarily (although sometimes for millions of years) trapped by the giant planets, similar to those seen for the terrestrial planets\cite{christou00,moraismorbidelli02,moraismorbidelli06}. 

\bibliographystyle{Science}
\bibliography{a13m}

\begin{thebibliography}{10}

\bibitem{SM}
See supplementary materials.

\bibitem{alexandersen13a}
M.~{Alexandersen}, J.~{Kavelaars}, J.~{Petit}, B.~{Gladman}, G.~V. {Williams},
  {\it Minor Planet Electronic Circulars\/} p.~19 (2013).

\bibitem{mikkola06}
S.~{Mikkola}, K.~{Innanen}, P.~{Wiegert}, M.~{Connors}, R.~{Brasser}, {\it
  Monthly Notices of the Royal Astronomical Society\/} {\bf 369}, 15 (2006).

\bibitem{chianglithwick05}
E.~I. {Chiang}, Y.~{Lithwick}, {\it Astrophysical Journal\/} {\bf 628}, 520
  (2005).

\bibitem{sheppardtrujillo10b}
S.~S. {Sheppard}, C.~A. {Trujillo}, {\it Astrophysical Journal Letters\/} {\bf
  723}, L233 (2010).

\bibitem{nesvornydones02}
D.~{Nesvorn{\'y}}, L.~{Dones}, {\it Icarus\/} {\bf 160}, 271 (2002).

\bibitem{dvorak10}
R.~{Dvorak}, {\'A}.~{Bazs{\'o}}, L.-Y. {Zhou}, {\it Celestial Mechanics and
  Dynamical Astronomy\/} {\bf 107}, 51 (2010).

\bibitem{kortenkamp04}
S.~J. {Kortenkamp}, R.~{Malhotra}, T.~{Michtchenko}, {\it Icarus\/} {\bf 167},
  347 (2004).

\bibitem{kortenkampjoseph11}
S.~J. {Kortenkamp}, E.~C.~S. {Joseph}, {\it Icarus\/} {\bf 215}, 669 (2011).

\bibitem{gladman08}
B.~{Gladman}, B.~G. {Marsden}, C.~{Vanlaerhoven}, {\it {Nomenclature in the
  Outer Solar System}\/} (The University of Arizona Press, 2008), pp. 43--57.

\bibitem{lykawkamukai07}
P.~S. {Lykawka}, T.~{Mukai}, {\it Icarus\/} {\bf 192}, 238 (2007).

\bibitem{dones99}
L.~{Dones}, {\it et~al.\/}, {\it Icarus\/} {\bf 142}, 509 (1999).

\bibitem{tiscarenomalhotra03}
M.~S. {Tiscareno}, R.~{Malhotra}, {\it Astronomical Journal\/} {\bf 126}, 3122
  (2003).

\bibitem{moraismorbidelli02}
M.~H.~M. {Morais}, A.~{Morbidelli}, {\it Icarus\/} {\bf 160}, 1 (2002).

\bibitem{moraismorbidelli06}
M.~H.~M. {Morais}, A.~{Morbidelli}, {\it Icarus\/} {\bf 185}, 29 (2006).

\bibitem{petit11}
J.-M. {Petit}, {\it et~al.\/}, {\it Astronomical Journal\/} {\bf 142}, 131
  (2011).

\bibitem{kaib11}
N.~A. {Kaib}, R.~{Ro{\v s}kar}, T.~{Quinn}, {\it Icarus\/} {\bf 215}, 491
  (2011).

\bibitem{hornerevans06}
J.~{Horner}, N.~{Wyn Evans}, {\it Monthly Notices of the Royal Astronomical
  Society\/} {\bf 367}, L20 (2006).

\bibitem{levisonduncan97}
H.~F. {Levison}, M.~J. {Duncan}, {\it Icarus\/} {\bf 127}, 13 (1997).

\bibitem{gladman12}
B.~{Gladman}, {\it et~al.\/}, {\it Astronomical Journal\/} {\bf 144}, 23
  (2012).

\bibitem{elliot05}
J.~L. {Elliot}, {\it et~al.\/}, {\it Astronomical Journal\/} {\bf 129}, 1117
  (2005).

\bibitem{shankman13}
C.~{Shankman}, B.~J. {Gladman}, N.~{Kaib}, J.~J. {Kavelaars}, J.~M. {Petit},
  {\it Astrophysical Journal Letters\/} {\bf 764}, L2 (2013).

\bibitem{christou00}
A.~A. {Christou}, {\it Icarus\/} {\bf 144}, 1 (2000).

\bibitem{mikkola04}
S.~{Mikkola}, R.~{Brasser}, P.~{Wiegert}, K.~{Innanen}, {\it Monthly Notices of
  the Royal Astronomical Society\/} {\bf 351}, L63 (2004).

\bibitem{wiegert98}
P.~A. {Wiegert}, K.~A. {Innanen}, S.~{Mikkola}, {\it Astronomical Journal\/}
  {\bf 115}, 2604 (1998).

\bibitem{connors02}
M.~{Connors}, {\it et~al.\/}, {\it Meteoritics and Planetary Science\/} {\bf
  37}, 1435 (2002).

\bibitem{connors04}
M.~{Connors}, {\it et~al.\/}, {\it Meteoritics and Planetary Science\/} {\bf
  39}, 1251 (2004).

\bibitem{connors11}
M.~{Connors}, P.~{Wiegert}, C.~{Veillet}, {\it Nature\/} {\bf 475}, 481 (2011).

\bibitem{christouasher11}
A.~A. {Christou}, D.~J. {Asher}, {\it Monthly Notices of the Royal Astronomical
  Society\/} {\bf 414}, 2965 (2011).

\bibitem{scholl05}
H.~{Scholl}, F.~{Marzari}, P.~{Tricarico}, {\it Icarus\/} {\bf 175}, 397
  (2005).

\bibitem{dlfmdlfm13b}
C.~{de la Fuente Marcos}, R.~{de la Fuente Marcos}, {\it Monthly Notices of the
  Royal Astronomical Society\/} {\bf 432}, L31 (2013).

\bibitem{levison97}
H.~F. {Levison}, E.~M. {Shoemaker}, C.~S. {Shoemaker}, {\it Nature\/} {\bf
  385}, 42 (1997).

\bibitem{karlsson04}
O.~{Karlsson}, {\it Astronomy and Astrophysics\/} {\bf 413}, 1153 (2004).

\bibitem{dlfmdlfm13a}
C.~{de la Fuente Marcos}, R.~{de la Fuente Marcos}, {\it Astronomy and
  Astrophysics\/} {\bf 551}, A114 (2013).

\bibitem{sheppardtrujillo06}
S.~S. {Sheppard}, C.~A. {Trujillo}, {\it Science\/} {\bf 313}, 511 (2006).

\bibitem{sheppardtrujillo10a}
S.~S. {Sheppard}, C.~A. {Trujillo}, {\it Science\/} {\bf 329}, 1304 (2010).

\bibitem{parker13}
A.~H. {Parker}, {\it et~al.\/}, {\it Astronomical Journal\/} {\bf 145}, 96
  (2013).

\bibitem{chiang03}
E.~I. {Chiang}, {\it et~al.\/}, {\it Astronomical Journal\/} {\bf 126}, 430
  (2003).

\bibitem{brasser04}
R.~{Brasser}, S.~{Mikkola}, T.-Y. {Huang}, P.~{Wiegert}, K.~{Innanen}, {\it
  Monthly Notices of the Royal Astronomical Society\/} {\bf 347}, 833 (2004).

\bibitem{hornerlykawka10a}
J.~{Horner}, P.~S. {Lykawka}, {\it Monthly Notices of the Royal Astronomical
  Society\/} {\bf 405}, 49 (2010).

\bibitem{horner12}
J.~{Horner}, P.~S. {Lykawka}, M.~T. {Bannister}, P.~{Francis}, {\it Monthly
  Notices of the Royal Astronomical Society\/} {\bf 422}, 2145 (2012).

\bibitem{guan12}
P.~{Guan}, L.-Y. {Zhou}, J.~{Li}, {\it Research in Astronomy and
  Astrophysics\/} {\bf 12}, 1549 (2012).

\bibitem{hornerlykawka12}
J.~{Horner}, P.~S. {Lykawka}, {\it Monthly Notices of the Royal Astronomical
  Society\/} {\bf 426}, 159 (2012).

\bibitem{dlfmdlfm12a}
C.~{de la Fuente Marcos}, R.~{de la Fuente Marcos}, {\it Astronomy and
  Astrophysics\/} {\bf 545}, L9 (2012).

\bibitem{dlfmdlfm12c}
C.~{de la Fuente Marcos}, R.~{de la Fuente Marcos}, {\it Astronomy and
  Astrophysics\/} {\bf 547}, L2 (2012).

\bibitem{levisonduncan94}
H.~F. {Levison}, M.~J. {Duncan}, {\it Icarus\/} {\bf 108}, 18 (1994).

\bibitem{bottke00}
W.~F. {Bottke}, R.~{Jedicke}, A.~{Morbidelli}, J.-M. {Petit}, B.~{Gladman},
  {\it Science\/} {\bf 288}, 2190 (2000).

\bibitem{volkmalhotra08}
K.~{Volk}, R.~{Malhotra}, {\it Astrophysical Journal\/} {\bf 687}, 714 (2008).

\bibitem{murraydermott99}
C.~D. {Murray}, S.~F. {Dermott}, {\it {Solar system dynamics}\/} (Cambridge
  University Press, 1999).

\bibitem{namouni99a}
F.~{Namouni}, {\it Icarus\/} {\bf 137}, 293 (1999).

\bibitem{namouni99b}
F.~{Namouni}, A.~A. {Christou}, C.~D. {Murray}, {\it Physical Review Letters\/}
  {\bf 83}, 2506 (1999).

\end{thebibliography}

\begin{scilastnote}
\item The data are available at the IAU Minor Planet Centre's online database under MPEC K13F19. 

We thank N. Kaib for making the scattering object model available to us. 

We thank S. Lawler, C. Shankman and N. Kaib for proof-reading and constructive comments. 

M. Alexandersen, S. Greenstreet and B. Gladman were supported by the National Sciences and Engineering Research Council of Canada.

This work is based on observations obtained at the Canada-France-Hawaii Telescope, operated by the National Research Council of Canada, the Institut National des Sciences de l'Univers of the Centre National de la Recherche Scientifique of France and the University of Hawaii. 
\end{scilastnote}

{\bf{Supplementary Materials}}\\
www.sciencemag.org\\
Supporting online text\\
Figs. S1, S2, S3, S4\\
References (24-51)

\clearpage

\begin{center}
\includegraphics[width=1.0\textwidth]{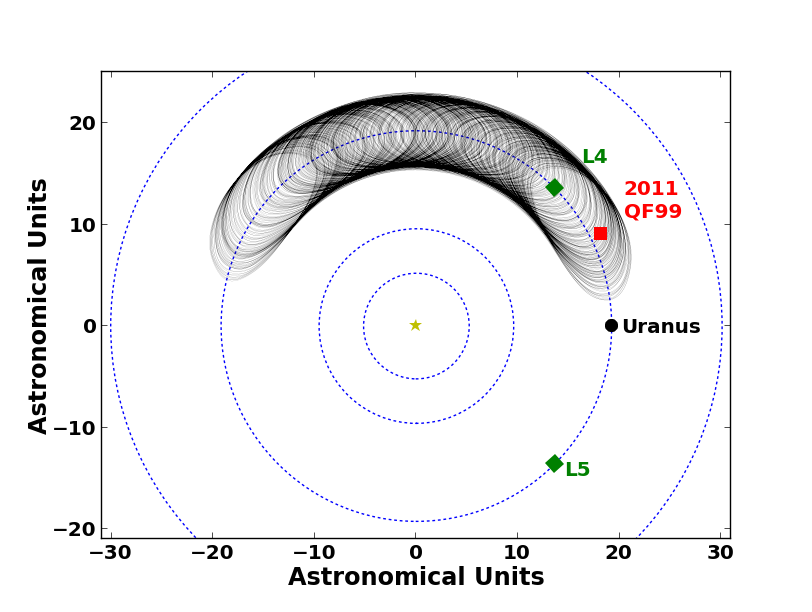}
\end{center}
\noindent {\bf Fig. 1.} 
The motion of 2011 QF$_{99}$. Shown here is the best fit trajectory of 2011 QF$_99$, from its current position (square) to $10$ libration periods ($59\unit{kyr}$) into the future.
The co-ordinate frame co-rotates with Uranus (on right) and dotted circles show the semi-major axis of the giant planets. 
Diamonds denote the L4 (upper) and L5 (lower) Lagrange points. 
The oval oscillations occur over one heliocentric orbit, while the angular extent around the Sun is the slower libration around L4. 

\clearpage

\begin{center}
\includegraphics[height=0.70\textheight]{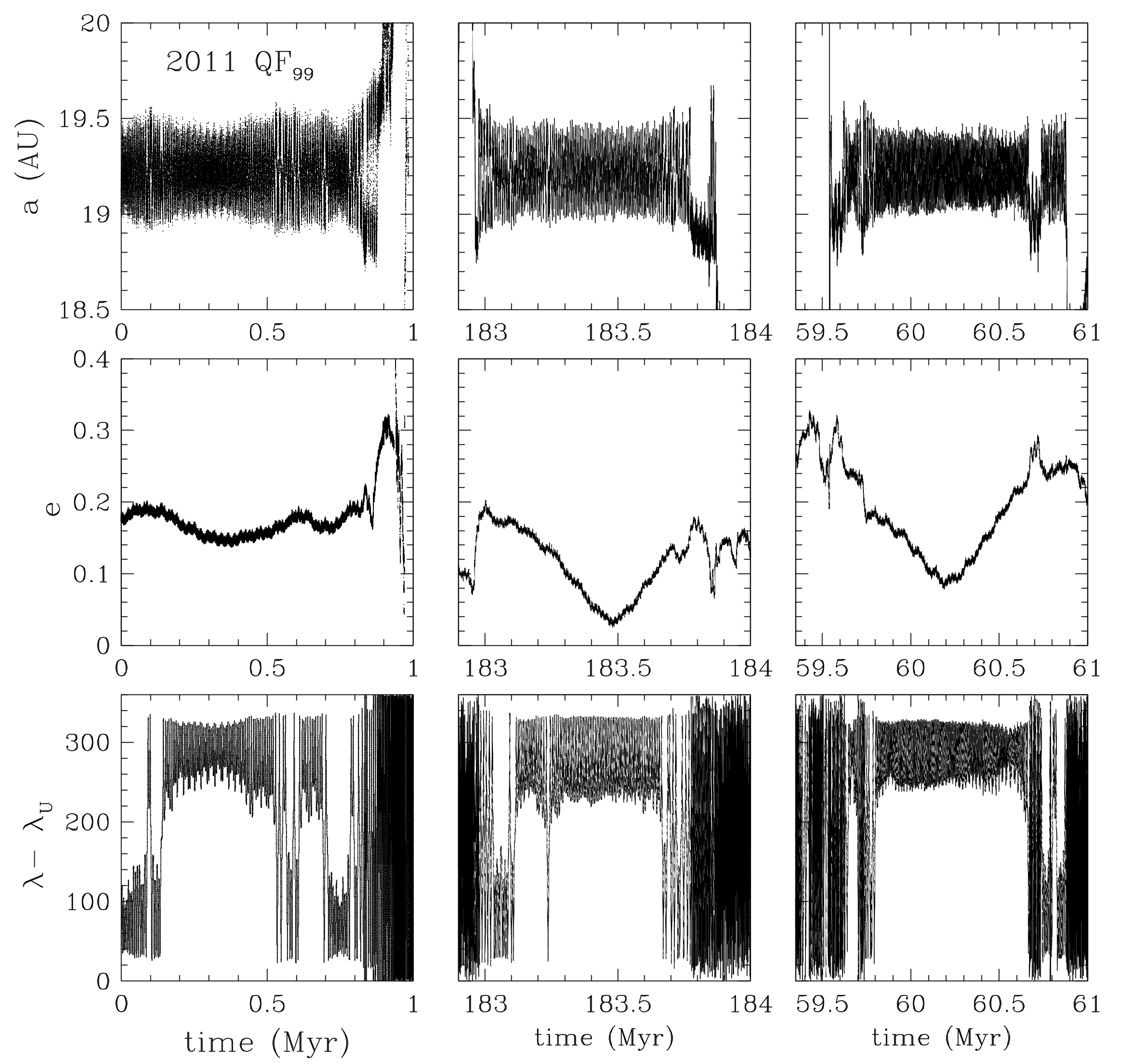}
\end{center}
\noindent {\bf Fig. 2.} Orbital evolution of temporary Uranian co-orbitals. 
Left column: Evolution of the nominal semi-major axis $a$, eccentricity $e$ and resonant angle $\lambda-\lambda_U$ of 2011 QF$_{99}$ for $1\unit{Myr}$ into the future. 
Centre and right columns: Evolution for two temporary Uranian co-orbitals from our dynamical simulations for intervals in which their evolution is similar to that of 2011 QF$_{99}$, showing that Centaurs can naturally become temporarily-trapped Uranian Trojans.
Times are from the initial condition for the $a_0>34\unit{AU}$ scattering orbit. 

\clearpage

\begin{center}
\includegraphics[height=0.52\textheight]{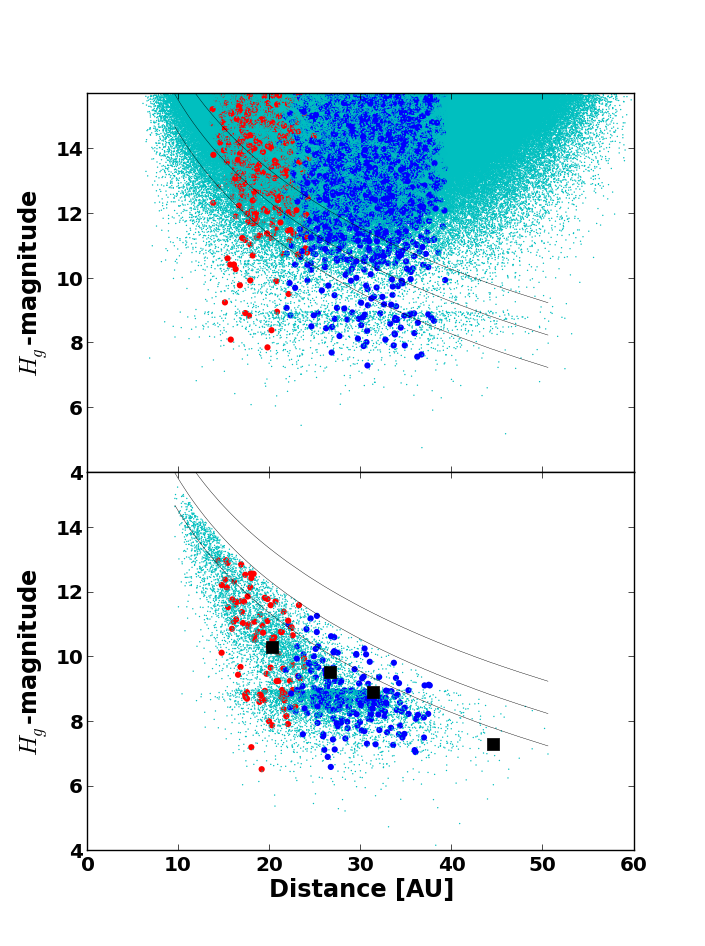}
\end{center}
\noindent {\bf Fig. 3.} Results of our survey simulations.
Top: $10^6$ objects drawn from the model population of scattering objects (tiny cyan dots), Uranian co-orbitals (red) and Neptunian co-orbitals (blue), used as the intrinsic population in our survey simulation. 
The populations have the relative fractions from the model, but the co-orbitals have larger symbols to enhance visibility.
Bottom: Objects detected by our survey simulator, using the same colour scheme. 
The large black squares are the four real $a<34\unit{AU}$ detections from our calibrated work, 2011 QF$_{99}$ being the upper-left one. 
Black curves correspond to apparent magnitudes $m_g=24.25$, $25.25$ and $26.25$, from bottom to top, which are roughly the survey limits of CFEPS\cite{petit11}, our new observations and a deep survey for Neptunian Trojans\cite{sheppardtrujillo10b}, respectively.
The effect of the divot $H$-mag distribution can be seen around $H_g=9.0$


\nocite{mikkola04}
\nocite{wiegert98}
\nocite{connors02}
\nocite{connors04}
\nocite{connors11}
\nocite{christouasher11}
\nocite{scholl05,dlfmdlfm13b}
\nocite{levison97}
\nocite{karlsson04}
\nocite{hornerevans06}
\nocite{dlfmdlfm13a}
\nocite{dvorak10}
\nocite{chianglithwick05}
\nocite{sheppardtrujillo06,sheppardtrujillo10a,parker13}
\nocite{chiang03} 
\nocite{brasser04,hornerlykawka10a,horner12,guan12}
\nocite{petit11,gladman12,hornerlykawka12}
\nocite{dlfmdlfm12a,dlfmdlfm12c}
\nocite{gladman12}
\nocite{gladman08}
\nocite{gladman08}
\nocite{kaib11}
\nocite{shankman13}
\nocite{levisonduncan94}
\nocite{bottke00}
\nocite{tiscarenomalhotra03,dones99,hornerevans06}
\nocite{levisonduncan97,volkmalhotra08}
\nocite{moraismorbidelli02}
\nocite{murraydermott99}
\nocite{hornerevans06}
\nocite{namouni99a,namouni99b,connors02,mikkola04,connors04,mikkola06,dlfmdlfm12a}
\nocite{dlfmdlfm12a}
\nocite{shankman13}
\nocite{shankman13}
\nocite{gladman12,hornerlykawka12}

\clearpage


\begin{center}
\includegraphics[width=1.0\textwidth]{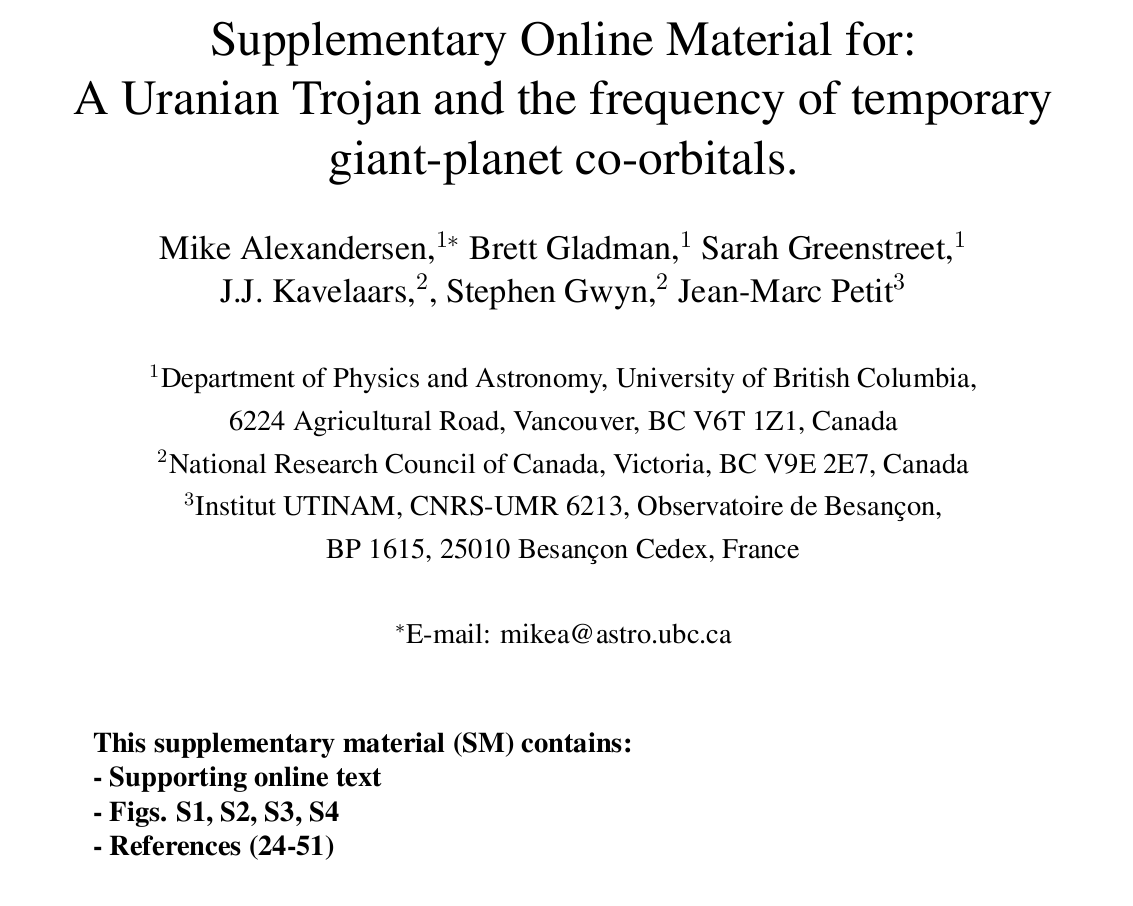}
\end{center}

\nocite{SM}
\nocite{alexandersen13a}
\nocite{mikkola06}
\nocite{chianglithwick05,sheppardtrujillo10b}
\nocite{nesvornydones02,dvorak10}
\nocite{kortenkamp04,kortenkampjoseph11}
\nocite{gladman08,lykawkamukai07}
\nocite{dones99,tiscarenomalhotra03}
\nocite{moraismorbidelli02}
\nocite{moraismorbidelli06}
\nocite{gladman08,petit11}
\nocite{kaib11}
\nocite{hornerevans06}
\nocite{levisonduncan97,tiscarenomalhotra03}
\nocite{moraismorbidelli02,moraismorbidelli06}
\nocite{gladman12}
\nocite{elliot05,petit11}
\nocite{sheppardtrujillo10b}
\nocite{sheppardtrujillo10b}
\nocite{shankman13}
\nocite{christou00,moraismorbidelli02,moraismorbidelli06}
\nocite{sheppardtrujillo10b}

\clearpage

\section*{Supplementary text}

\subsection*{Known Co-orbitals of the planets}

There are objects known to be in co-orbital motion with several of the planets in the Solar System, both as long-term stable, presumably primordial (by which we mean $\sim4\unit{Gyr}$ lifetimes) populations and also as temporary captures. 
Working outwards from the Sun:\\
{\bf{Venus}} has a temporary quasi-satellite\cite{mikkola04}.\\
{\bf{Earth}} has multiple unstable co-orbital companions. 
3753 Cruithne\cite{wiegert98} is on a complex orbit, a combination of horseshoe-quasi-satellite, due to its substantial inclination and eccentricity. 2002 AA$_{29}$\cite{connors02} exhibits periods of both temporary horseshoe and quasi-satellite behaviours. 
2003 YN$_{107}$\cite{connors04} is currently a quasi-satellite while 2010 TK$_7$\cite{connors11} is a temporary L4 Trojan. 
The most stable (longest duration of resonance occupation) known co-orbital of the Earth is 2010 SO16\cite{christouasher11}, which remains in a horseshoe orbit for more than $100\unit{kyr}$.\\
{\bf{Mars}} has eight known Trojans, all of which have been shown to be stable on at least Gyr time scales\cite{scholl05,dlfmdlfm13b}.\\
{\bf{Jupiter}} has almost 6000 known long-term stable Trojan asteroids (MPC database, 2013 June 11) in its Trojan clouds, which are believed to outnumber the main asteroid belt\cite{levison97}. In terms of temporary co-orbitals, only a few very short-term, $<1\unit{kyr}$, captures have been identified\cite{karlsson04}.\\
{\bf{Saturn}} does not have any known co-orbitals and its co-orbital phase-space is believed to be highly unstable\cite{nesvornydones02}. Orbital simulations\cite{hornerevans06} show temporary captures are possible. \\
{\bf{Uranus}} has one known temporary ($\sim20\unit{kyr}$) horseshoe companion\cite{dlfmdlfm13a} and 2011 QF$_{99}$, the temporary L4 Trojan reported here. 
The Uranian co-orbital region is thought to be mostly unstable\cite{nesvornydones02} although some stable niches may exist\cite{dvorak10}.\\
{\bf{Neptune}} was recently discovered to have a large stable Trojan population which might outnumber the Jovian Trojans\cite{chianglithwick05}. 
Neptune currently has nine known Trojans, of which 6 are known to be stable over the age of the Solar System\cite{sheppardtrujillo06,sheppardtrujillo10a,parker13}. 
The first discovered Neptunian Trojan, 2001 QR$_{322}$\cite{chiang03}, as well as 2008 LC$_{18}$, have orbital uncertainties straddling the boundary between long-term stable and temporary librators and may be short-lived\cite{brasser04,hornerlykawka10a,horner12,guan12} although primordial orbits are also possible. 
The Canada-France Ecliptic Plane Survey (CFEPS) discovered the first known Neptunian Trojan that is certainly unstable on a short time scale (2004 KV$_{18}$, Fig. S2 left column)\cite{petit11,gladman12,hornerlykawka12}. 
Recently others\cite{dlfmdlfm12a,dlfmdlfm12c} have run short numerical integrations of known Centaurs and identified several temporary Neptunian co-orbitals: a temporary quasi-satellite, three temporary Trojans and a temporary horseshoe. 
Although some of these classifications are still insecure, the number of known transient Neptunian co-orbitals is now, maybe surprisingly, of order the same as the long-term stable co-orbitals. \\

\subsection*{Observations}
 
  Our study used the Canada-France-Hawaii Telescope to cover a $20\unit{sq.deg.}$ patch of sky on the ecliptic near RA=$2\unit{hr}$, Dec=$15\degree$, in the $r$-band filter. 
  This direction was selected to optimise the search for resonant objects, as RA=$2\unit{hr}$ is $\approx60\degree$ ahead of Neptune, the region in which L4 Trojans reside, as well as where n:2 and n:3 resonant objects come to pericentre\cite{gladman12}. 
  The searched fields were also near the Uranian L4 point (Fig. 1).
  Discovery observations were obtained in October 2011 and tracking observations were made throughout 2011 and 2012, succeeding in a $100\%$ tracking fraction for $m_r<24.55$. 
  The detection efficiency was measured by implanting fake objects into the observations. 
  The search had a detection efficiency of $88\%$ for objects brighter than $m_r=23.0$ with a magnitude limit at $m_r=24.55$. 
  Here, the magnitude limit is quoted as the magnitude at which the detection efficiency has dropped to half of the maximum efficiency. 
  The observations were searched for objects moving at rates from $0.4\unit{"/hr}$ to $10.3\unit{"/hr}$, corresponding to distances from over $300\unit{AU}$ to $\sim10\unit{AU}$, respectively. 

\subsection*{Details on classification of 2011 QF$_{99}$}

After a short integration, which showed that 2011 QF$_{99}$ was librating
around the Uranian L4 point, we ran a $10\unit{Myr}$ classification using the Solar System Beyond Neptune classification algorithm\cite{gladman08}.
In that algorithm, instead of using a Gaussian covariance matrix based
on the rms scatter of the observations, a Monte-Carlo technique identifies a slew of orbits compatible with the astrometric residual pattern. Of these, the orbits with the the `maximal' and `minimal' semi-major axis, as well as the best fit orbit, are used for classification. 
In this case the three heliocentric semi-major axes are $19.167\unit{AU}$, $19.175\unit{AU}$, and $19.183\unit{AU}$
(these are not Gaussian errors, but should rather be interpreted as
the maximum allowable range).
These three orbits are then integrated for $10\unit{Myr}$ into the future to look for resonant behavior (Fig. S1).
For 2011 QF$_{99}$, all 3 orbits escape the co-orbital 1:1 resonance in less
than $3\unit{Myr}$, so 2011 QF$_{99}$ is not classified by this algorithm as a stable
resonant object; instead it is only temporarily (by astronomical
standards) in the co-orbital state.
Fig. S1 shows the three evolutions for the next $1\unit{Myr}$ into the
future.
As shown (and also true in all other experiments we have conducted
with different time steps), 2011 QF$_{99}$ always executes 10 or more
angular librations of the $\phi_{11}$ resonant argument around
the L4 point, over the next $70\unit{kyr}$ or more.
Due to the chaotic nature of the resonance dynamics, by this point
the orbits have sufficiently separated that they show different
evolutions as to when they exit the L4 libration.
This exit is usually into another form of co-orbital behaviour
(transition to an L5 librator or a horseshoe orbit).
We numerically estimate that the object's current Lyapunov time
scale (for chaotic divergence) is $\sim10\unit{kyr}$ while co-orbital.
As is usual for such a chaotic trajectory, there is a
range of times to exit the resonance; we find that the co-orbital status of 2011 QF$_{99}$ is very likely to be maintained for the next half
million years into the past or future.

On time scales of $\sim$Myr all orbits we have integrated 
(both those within tens of sigmas from the best fit orbit, as in the Gladman et al.\cite{gladman08} algorithm, and those orbits that could plausibly be caused by systematic errors, as described in the main text)
leave the resonance to rejoin the Centaur population from which the object must have come.

\subsection*{Details on dynamical integrations}

{\bf{The dynamical integrations}} computed to model the steady-state 
distribution of scattering objects in the $a<34\unit{AU}$ region for this
work were set up using a subset of particles from the Kaib et al. 
(2011 - KRQ11 henceforth)\cite{kaib11} TNO population model. 
We use the term ``steady-state'' only to refer to a constant relative distribution of objects (ie. the distribution of Centaurs being constant, and thus the co-orbital fraction being constant), not to denote a constant absolute population, as the population of Scattering Disk objects is depleting on Gyr time scales. 
This subset consisted of 17,800 particles with initial semi-major axis $34\unit{AU}<a<200\unit{AU}$ and 
which had their semi-major axes deviate by more than $1.5\unit{AU}$ during the last $10\unit{Myr}$ of the KRQ11 model integrations. 
This populations of scattering TNOs were used as the initial conditions for the orbital integrations in this work.
Two different KRQ11 models were used independently: one generated from a primordial inclination distribution that was dynamically cold when the particles left the giant planet region $4\unit{Gyr}$ earlier (the ``cold'' model)\cite{kaib11}, the other using an initially hot distribution (the ``hot'' model) integrated in the same way as the KRQ11 model\cite{shankman13}. 

To perform this computation, we used the N-body code SWIFT-RMVS4 (provided by Hal Levison, based on the original SWIFT\cite{levisonduncan94}). 
A base time step of $73\unit{days}$ was used and the orbital elements were output every $300\unit{years}$ for any particle which at that moment had $a<34\unit{AU}$. 
The gravitational influences of the four giant planets were included. 
Particles were removed from the simulation when they hit a planet, went outside $2000\unit{AU}$ or inside $6\unit{AU}$ from the Sun (resulting in rapid removal from the Solar System by Jupiter), or the final integration time of $1\unit{Gyr}$ was reached. 

The goal of these orbital integrations was twofold: to search for temporary co-orbital trapping and to construct 
the steady-state orbital distribution of scattering TNOs which reach the 
giant planet region, chosen to be the $a<34\unit{AU}$ region. 
The steady-state orbital distribution is expressed as a grid with $a<34\unit{AU}$, $e<1.0$, $i<180\degree$, with cells of volume $0.5\unit{AU}\times0.02\times2.\degr0$. 
The cumulative time spent by all particles in each cell is normalized 
to the total time spent by all particles in all cells in the $a<34\unit{AU}$ 
region. 
This illustrates the steady-state distribution of objects in the $a<34\unit{AU}$ region (see Fig. S3) supplied by the scattering TNO population. 
This ``residence time'' probability distribution\cite{bottke00} can be interpreted as the steady-state fraction of scattering TNOs in each cell. Fig. S3 
shows two projections of the residence time probability 
distributions of the $a<34\unit{AU}$ region for the two KRQ11 
population models. 
From these plots it is clear the scattering 
TNO population enters the giant planet region ($a<34\unit{AU}$) at moderate eccentricities and inclinations. 
Although the hot model does produce higher inclinations, it is clear from Fig. S3, and is confirmed by our survey simulator, that the choice of input model does not make a large difference for our results. 
We therefore only describe results from the simulations using the KRQ11 ``hot'' model in the main text and from here on (unless otherwise noted). 

To confirm that our Centaur distribution is in fact in steady-state (fractionally, not absolutely), we divide our $1\unit{Gyr}$ integrations into $<100\unit{Myr}$ and $100-1000\unit{Myr}$ intervals. 
The $<100\unit{Myr}$ interval contains about half (cold model) or one-third (hot model) of all entries into the $a<34\unit{AU}$ regime. 
In all four cases the fraction of temporary co-orbitals is the same ($0.31-0.62\%$ for Uranus and $2.3-3.3\%$ for Neptune) to well within a factor of 2 accuracy and the distribution of Centaurs are all similar to those seen in Fig. S3. 
We thus believe this justifies treating the relative distribution of objects as being time independent, despite the absolute scattering/Centaur population slowly decreasing. 

This is not the first work of its kind to perform numerical integrations in order to construct the steady-state population of Centaurs with $a<34\unit{AU}$ from a scattering TNO population. 
Some works present numerical integrations of Centaurs (both known and test populations)\cite{tiscarenomalhotra03,dones99,hornerevans06} initially within the giant planet region. 
Those that have modelled the evolution of scattering TNOs into the $a<34\unit{AU}$ Centaur region\cite{levisonduncan97,volkmalhotra08} did not search for temporary (often $<100\unit{kyr}$) co-orbital captures. 

Work similar to that presented here has been performed simulating near Earth asteroids captured as temporary co-orbitals of Earth\cite{moraismorbidelli02}. 
Those authors found that the Earth's temporary co-orbitals often experience several 
co-orbital phases, each of average duration 25 kyr (none longer than 1 Myr).

{\bf{Co-orbital detection}}.
To diagnose whether particles are co-orbital, the orbital history (at $300\unit{year}$ output intervals) was scanned using a running window $30\unit{kyr}$ long. 
This window-size was chosen, as this is several times longer than the typical Trojan libration period. 
While the formal definition of co-orbital is that the resonant angle 
$\phi_{11}=\lambda-\lambda_{Planet}$ librates, detecting this is difficult to automate. 
As an automatic process is necessary to filter the large amounts of output from our dynamical simulations ($110\unit{GB}$), 
we used a simpler algorithm which we believe diagnoses co-orbitals well.
A particle was classified as a co-orbital if, within the running 
window, both its average semi-major axis was less than 0.2 AU from 
the average semi-major axis of a given planet and no individual 
semi-major axis value deviated more than $R_H$ from that of the planet. 
Here $R_H$ is the planet's Hill-sphere radius\cite{murraydermott99}, where $R_H=0.47\unit{AU}$ for Uranus and $R_H=0.77\unit{AU}$ for Neptune. 
If these criteria were met, the orbital elements 
and current integration time for that particle (at the time-centre of the window) were output to indicate co-orbital motion in that window. 
The window centre then advances by a 
single 300-year output interval and performs the diagnosis determination again. 
In this manner, consecutive identifications of a particle in co-orbital 
motion with a planet can be recorded as a single ``trap'' until 
the object is scattered away. A minor shortcoming of this method of 
co-orbital identification is that the beginning and end of each 
trap is not diagnosed well due to the ends of the window not falling 
entirely within the trap at these times. This method 
provides us with estimates of the duration of traps, each of which must be greater than $30\unit{kyr}$ to be diagnosed by this analysis.

Numerical integrations of known Centaurs have previously been performed\cite{hornerevans06} in order to study 
the capture of Centaurs as temporary co-orbitals of the giant planets.
That study found that captures are generally short ($10$-$100\unit{kyr}$, none greater than $500\unit{kyr}$) with $0.29\%$ of 23,328 Centaur clones experiencing a co-orbital capture during a 3 Myr simulation, of which $15\%$, $80\%$, $5\%$ and $0\%$ of these captures at Jupiter, Saturn, Uranus and Neptune, respectively. 
The previous study used clones of the known Centaurs for initial conditions, a starting condition heavily biased towards smaller semi-major axes. 

In the work presented here, where Centaurs are provided from a bias-free external Scattering Disk, we find that the average length of captures in co-orbital motion with Uranus is $114\unit{kyr}$ and with Neptune is $84\unit{kyr}$. 
We were surprised to find that objects that experience at least one episode of co-orbital capture have a median of 2 captures with Uranus or 6 captures with Neptune. 
Objects typically escape with low relative velocities, so multiple temporary captures is not surprising. 
Some objects experience temporary co-orbital captures with both planets (see Fig. S2, right column). 
Due to the $r_{min}=6\unit{AU}$ distance cut in the integrations, which removes high-eccentricity Saturn crossing Centaurs before they potentially could get trapped into co-rotation, we did not reliably measure the Saturnian trapping fraction, but estimate it at $\ll0.1\%$ of the incoming scattering population.

{\bf{Resonant island classification}}. For each time step a particle has been
deemed co-orbital, we wish to determine in which of the four
resonant islands the particle is librating, i.e. whether it is a horseshoe,
L4 Trojan, L5 Trojan, or quasi-satellite. 
As our co-orbital detection algorithm (described above) produces $\sim25,000$ trapping episodes, these cannot be inspected manually and require another automated process. 
As for the detection algorithm, this is also hard to automate, especially because complex variations and combinations can exist for high inclinations. 
For our classification algorithm, we examine the behavior
of the resonant angle ($\phi_{11}=\lambda-\lambda_{p}$) in each $30\unit{kyr}$ window.
If $\phi_{11}$ remained in the leading or trailing hemisphere during a window, we assigned the particle to the L4 or L5 state (respectively). 
If $\phi_{11}$ crosses $180\degree$ at any time during the interval, 
then it was labelled a horseshoe orbit. 
The remaining objects are assumed to be quasi-satellites, as they must be co-orbitals that cross between leading and trailing at $\phi_{11}=0\degree$ and not at $180\degree$.
The possibility of erroneous classification exists, however in a manually inspected subset we find these errors affect $\ll10\%$ of cases, thus not affecting our co-orbital fraction and resonant island distribution estimates greatly, supporting our goal of better than factor of two accuracy. 

\subsection*{Quasi-satellites}

Quasi-satellites make up $10\%$ of the steady-state Uranian and Neptunian co-orbitals in our numerical integrations. 
This is thus a rare state, but of great dynamical interest\cite{namouni99a,namouni99b,connors02,mikkola04,connors04,mikkola06,dlfmdlfm12a}.
There is currently one known temporary quasi-satellite of Neptune\cite{dlfmdlfm12a}.
The current existence of one known temporary quasi-satellite, out of a total of $\sim6$ known temporary Uranian and Neptunian co-orbitals, fits into our general picture of temporary traps in co-orbital states. 
Fig. S4 depicts the semi-major axis, eccentricity, and resonant angle evolution of two temporary quasi-satellite captures found in our numerical integrations. 
The capture shown on the left is a quasi-satellite with Neptune for a duration of 694.5 kyr before it scatters away. 
The capture on the right in Fig. S4 remains a quasi-satellite with Uranus for 1.45 Myr before leaving the co-orbital state.

\subsection*{Survey Simulator}

The CFEPS Survey Simulator works by either drawing random objects from a ready-made list of objects or from a model orbital distribution within the simulator. 
The object is then allocated an $H$-magnitude, drawn from a model $H$-mag distribution. 
The simulator then proceeds to test whether or not a given object would be detectable by any of the surveys that it has been set up to simulate, taking into account the position of the observed fields, the position of the object, the detection efficiency as a function of object magnitude, as well as rate and direction of motion of the object. 
We only use our current campaign and CFEPS, as those are the only observations for which we have access to all the required information needed to simulate the survey. 
For this paper, we draw scattering objects with $a<34\unit{AU}$ from the output of the dynamical simulation.
However, the dynamical simulation only generated a few hundred co-orbital objects and their distribution was therefore insufficiently sampled to be used directly as a source for the survey simulator. 
Instead we drew co-orbitals from a model distribution set up to resemble the co-orbital distribution found in the dynamical simulation, with relative fractions in each population (Centaurs, Uranian Trojans, Neptunian horseshoes, etc.) set to be the same as that found from the dynamical simulation. 

\subsection*{$H$-mag distribution}

Minor body populations can typically be well represented, over some range of magnitudes, by an exponential absolute magnitude, $H$, distribution of the form 
$$
\frac{\mathrm{d}N}{\mathrm{d}H}\propto10^{\alpha H}
$$
where $\mathrm{d}N/\mathrm{d}H$ is the number of objects per $H$-mag. 
The cumulative $H$-mag distribution can be described by integrating this to get
$$
N(<H)\propto10^{\alpha H}
$$
where $N(<H)$ is the cumulative number of objects brighter than $H$. 
Note that both distributions have the same exponent. 

As described in the main text, a single exponential $H$-mag distribution does not agree with our observations; 
the single exponential distribution predicts that small ($H_g>11.0$) objects should completely dominate, accounting for $81\%$ of the simulated detections, whereas our observations have no such objects. 
The probability of 0 small objects out of 4 detections given this expectation is $0.12\%$, rejecting the single exponential distribution at $>99\%$ confidence.
Previous work\cite{shankman13} has shown that the Scattering Disk objects cannot follow a single exponential function and is better modelled by a divot distribution. 
The divot $H$-mag distribution is simply two different distributions of the form above, one relevant for magnitudes brighter than the divot, the other relevant for magnitudes fainter than the divot, with a discontinuity at the divot.
As we are pursuing the notion that the temporary co-orbitals are captured Centaurs, which themselves are Scattering Disk objects scattered inwards, it is to be expected that the temporary co-orbitals and Centaurs follow the same $H$-mag distribution as the Scattering Disk. 
Adopting the divot $H$-mag distribution produces results in better agreement with our observations (see Fig. 3); small ($H_g>11.0$) objects now only provide $23\%$ of the detections. 
The probability of detecting 0 small objects out of four given this expectation is $36\%$, so this distribution is not in conflict with the observations and we adopt it. 
As the values of the contrast and post-divot slope are not uniquely determined\cite{shankman13}, we explored reasonable ranges but found less than factor of two changes in the detected fraction of co-orbitals.

To further illustrate the improvement in using a divot $H$-mag distribution over the single exponential distribution, 
we test our four detections against the simulated detections using the Kolmogorov-Smirnov (KS) and Anderson-Darling (AD) tests. 
Using a single exponential $H$-mag distribution, the KS test gives $0.01\%$ and $0.15\%$ probability that our four detections were drawn from the same $H$-mag and distance distributions as the simulated detections, respectively. 
The AD test gave $0.03\%$ and $0.02\%$ for those same parameters. 
Using the divot $H$-mag distribution, the KS test gives $67\%$ and $27\%$ probability that our four detections were drawn from the same $H$-mag and distance distributions as the simulated detections (see Fig. 3, bottom), respectively. 
The AD test gave $64\%$ and $7.4\%$ for those same parameters.
Thus the single exponential is rejectable at greater than $99\%$ confidence for all four tests, while the divot distribution is not rejectable at $95\%$ confidence by any of the tests. 

\clearpage

\begin{center}
\includegraphics[height=0.60\textheight]{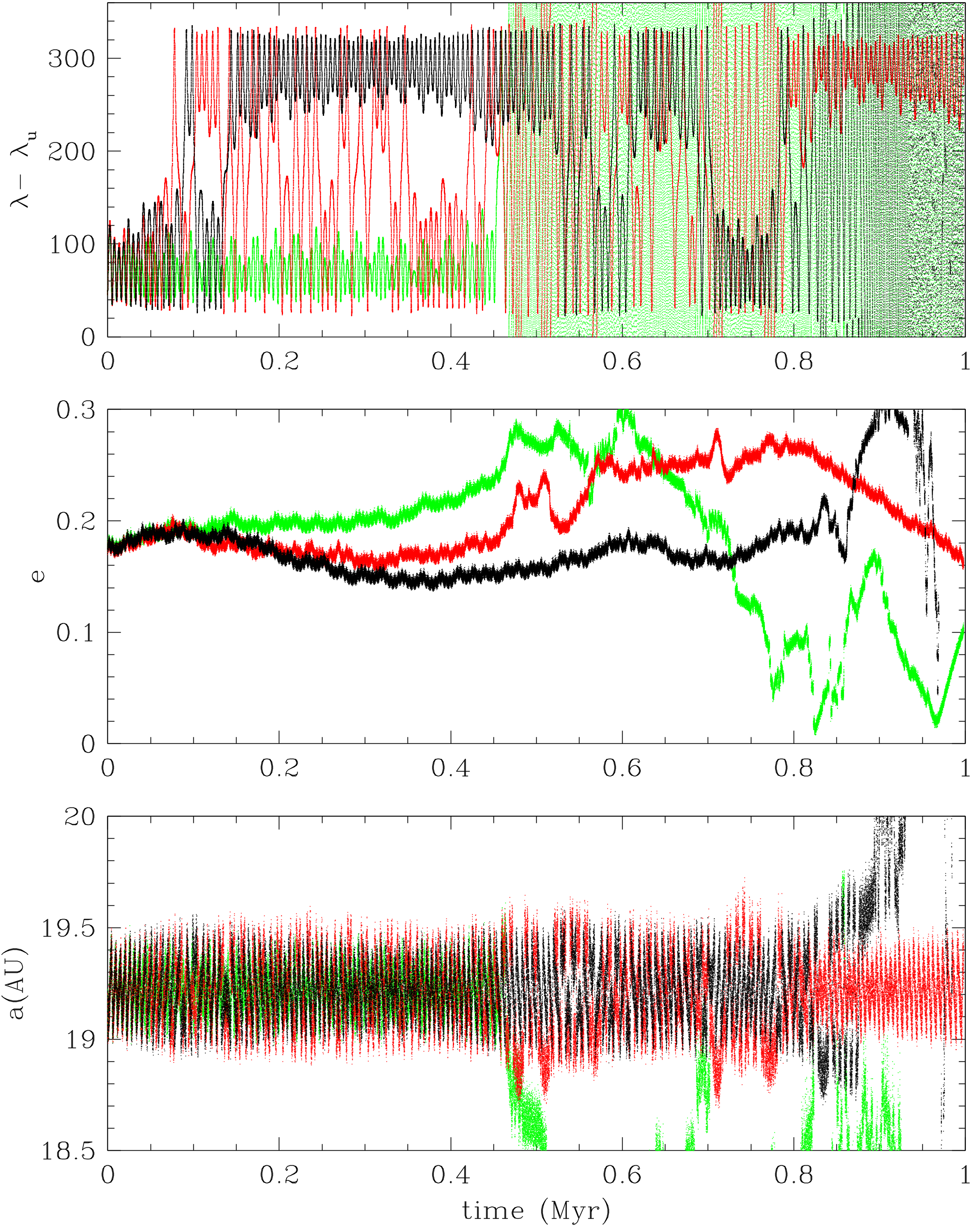}
\end{center}
\noindent {\bf Fig. S1.} The future evolution of 2011 QF$_{99}$. 
Shown is the numerical integration of the best-fit orbit (black) as well as the smallest and largest semi-major axis orbits compatible with the astrometry (green and red, respectively). 
The detailed evolution is highly chaotic, but all three clones remain as L4 Trojans for at least $70\unit{kyr}$ and remain co-orbital for at least $450\unit{kyr}$.
The nominal orbit (black) switches back and forth from L4 to L5, before a horseshoe period at $t\approx+800\unit{kyr}$ and then escape from the resonance at $t\approx+880\unit{kyr}$. 
The small-$a$ clone (green) remains around L4 the longest, until breaking loose around $t\approx+450\unit{kyr}$ and never getting recaptured. 
The large-$a$ clone (red) is the most stable and remains in or near the co-orbital state for the entire $1\unit{Myr}$ shown here, before leaving soon after (not shown). 

\clearpage

\begin{center}
\includegraphics[height=0.70\textheight]{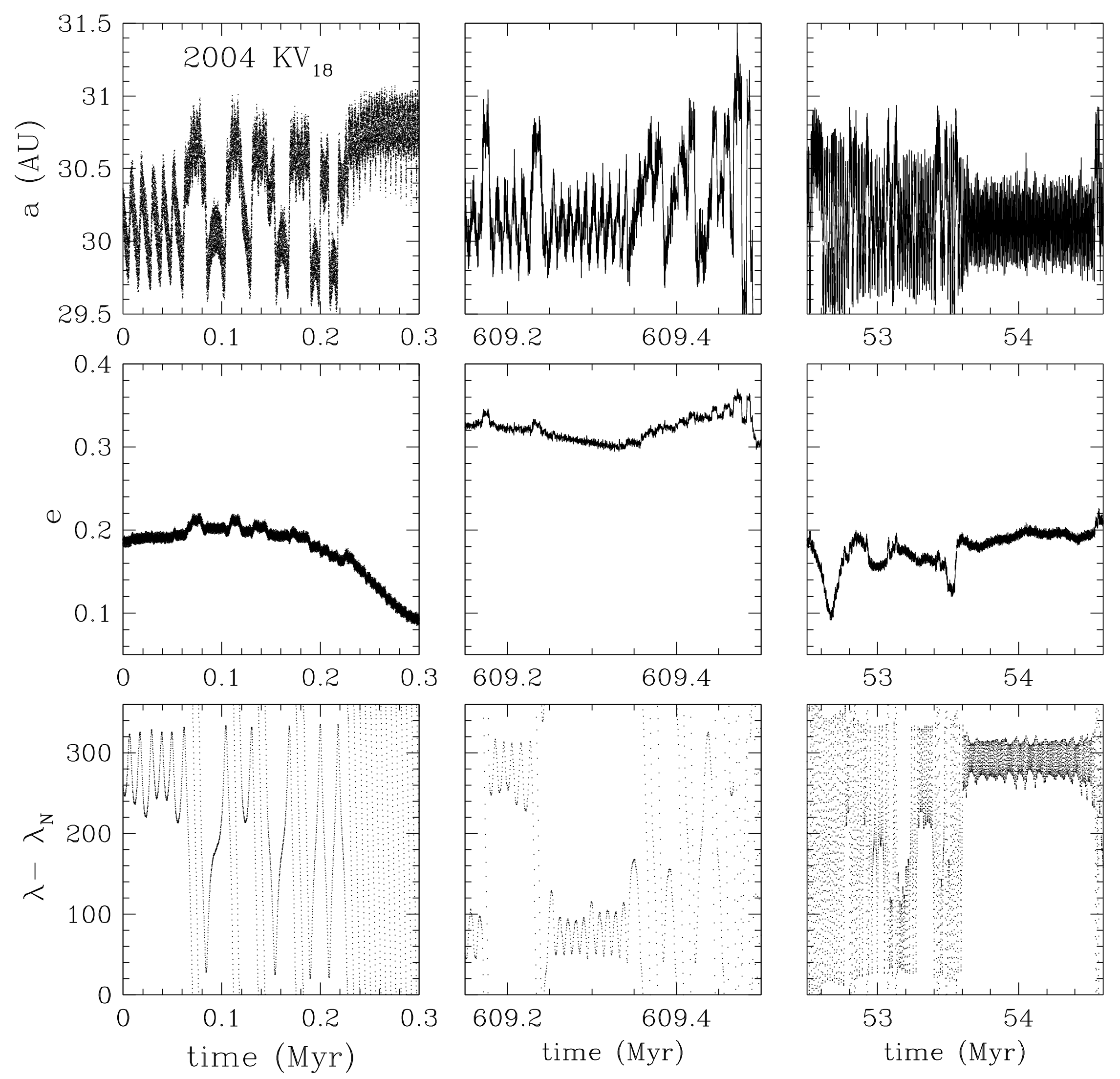}
\end{center}
\noindent {\bf Fig. S2.} Orbital evolution of temporary Neptunian co-orbitals. 
Left column: Evolution (for $0.3\unit{Myr}$ into the future) of the semi-major axis $a$, eccentricity $e$ and resonant angle $\lambda-\lambda_N$ of 2004 KV$_{18}$ (the certainly-unstable Neptunian Trojan\cite{gladman12,hornerlykawka12}).
Centre and right columns: Evolution for two temporary Neptunian co-orbitals from our dynamical simulations for intervals in which their evolution is similar to that of 2004 KV$_{18}$. 
Note: The object on the right is the same object as in right column of Fig. 2. This object experiences co-orbital motion with both Uranus and Neptune, with $\sim5\unit{Myr}$ between the two temporary captures. 

\clearpage

\begin{center}
\includegraphics[width=1.0\textwidth]{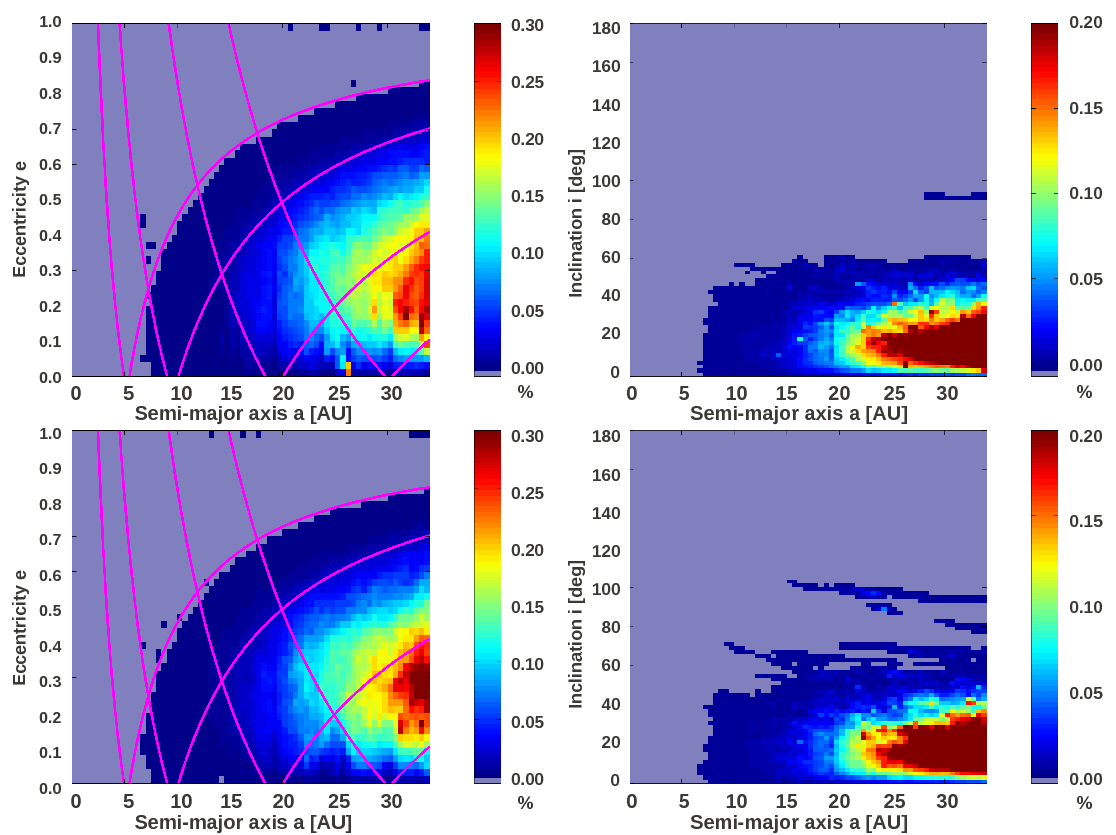}
\end{center}
\noindent {\bf Fig. S3.}
Residence time probability distributions. 
The top and bottom plots show the distribution resulting from the initially cold and hot KRQ11 model, respectively; the two different initial populations clearly produce very similar $a<34\unit{AU}$ steady-states. 
To monitor the orbital 
evolution of each particle, a grid of $a$, $e$, $i$ cells was 
placed throughout the giant planet region from $a<34\unit{AU}$, $e<1.0$, 
and $i<180\degree$ with volume $0.5\unit{AU}\times0.02\times2.\degr0$. 
The $a$, $e$ plot is summed over $i$, and the $a$, $i$ plot is summed over $e$. 
The color scheme 
represents the percentage of the steady-state Centaur population 
contained in each bin; Red colors represent cells where there is 
a high probability of particles spending their time. The curves indicate Jupiter, Saturn, Uranus, and Neptune crossing 
orbits. 

\clearpage

\begin{center}
\includegraphics[width=1.0\textwidth]{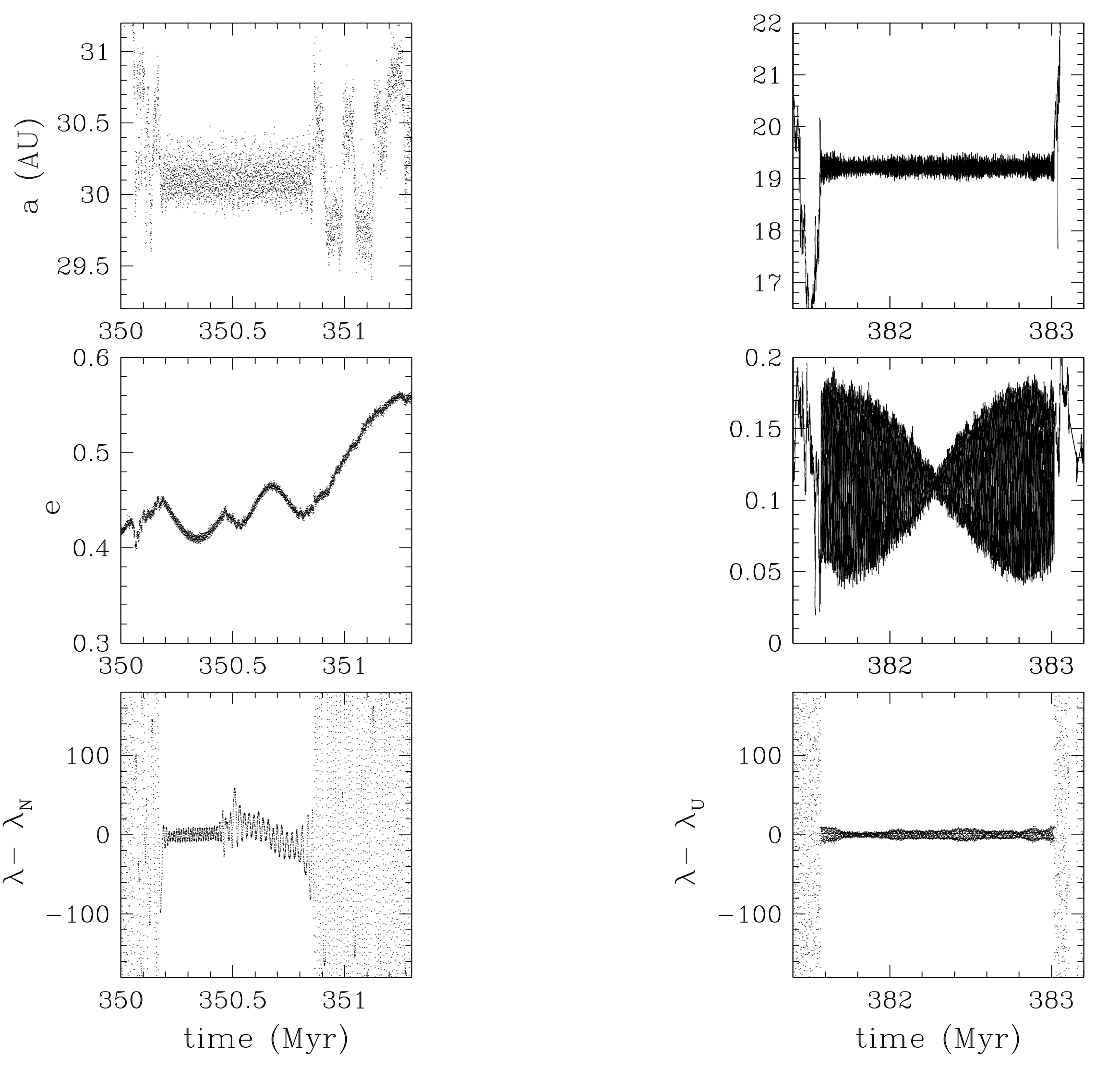}
\end{center}
\noindent {\bf Fig. S4.} Orbital evolution of temporary quasi-satellites. 
Evolution of the semi-major axis $a$, eccentricity $e$ and resonant angle $\lambda-\lambda_{Planet}$ for two temporary quasi-satellites captures from our dynamical simulation, one Uranian (right) and one Neptunian (left). 
Note the different scales.

\end{document}